\documentclass[a4paper,
               keeplastbox,   % flushend option: not to un-indent last line in References
               ]{jacow}
\usepackage{pdfpages,multirow,ragged2e} %
\makeatletter%
	\ifboolexpr{bool{xetex}}
	 {\renewcommand{\Gin@extensions}{.pdf,%
	                    .png,.jpg,.bmp,.pict,.tif,.psd,.mac,.sga,.tga,.gif,%
	                    .eps,.ps,%
	                    }}{}
\makeatother

\ifboolexpr{bool{xetex} or bool{luatex}} % test for XeTeX/LuaTeX
 {}                                      % input encoding is utf8 by default
 {\usepackage[utf8]{inputenc}}           % switch to utf8

\usepackage[USenglish]{babel}
\usepackage{float}
\ifboolexpr{bool{jacowbiblatex}}%
 {%
  \addbibresource{jacow-test.bib}
  \addbibresource{biblatex-examples.bib}
 }{}
\begin{document}

\title{Status of the LLRF system for SARAF project phase \uppercase\expandafter{\romannumeral2}}

\author{L. ZHAO \thanks{lu.zhao@cea.fr}, G. FERRAND, F. GOUGNAUD, R. DUPERRIER, N. PICHOFF and C. MARCHAND.\\ DACM/Irfu/CEA-Saclay, Gif-sur-Yvette, France }
	
\maketitle

\begin{abstract}
CEA is committed to the design, construction and commissioning of a Medium Energy Beam Transfer line and a  superconducting linac (SCL)  for SARAF accelerator in order to accelerate 5mA beam of either protons from 1.3 MeV to 35 MeV or deuterons from 2.6 MeV to 40 MeV. The Low Level RF (LLRF) is a subsystem of the CEA control domain for the SARAF-LINAC instrumentation. The top level requirement of the LLRF system has been presented in the last LLRF conference. The paper shows a simulink model to analyse and determinate the LLRF technical specification. The public bidding for SARAF LLRF is in the last phase: discussion with the selected company. The first prototype test will be performed at the start of 2020.
\end{abstract}
%==========================================
\section{Introduction}
The LLRF control system is the key component to regulate RF field in the cavity. It is used as a closed loop controller which maintains cavity gradient and phase stability when operating the cavity with beam. Digital and analog control systems have been developed at any number of institutions for control of RF cavities for both CW and pulsed operations in charged particle accelerators. The LLRF channels required in the frame of the SARAF-LINAC phase  \uppercase\expandafter{\romannumeral2}\cite{saraf2018} contract shall drive normal conducting cavities and superconducting (SC) cavities in CW mode only.

%==========================================
\section{Cavity Specification}
The SARAF-Linac Phase \uppercase\expandafter{\romannumeral2} will consist in 4 cryomodules with HWR cavities at the frequency of 176 MHz. The low-beta cavities are optimized to  $\beta_{opt} = 0.09$ and the high-beta cavities are optimized to $\beta_{opt} = 0.18$. Table \ref{tab:SCcavities} presents the critical parameters for the low- and high-beta cavities \cite{cavity}.  The frequency target of these cavities is 176.000 MHz in nominal operations.  The maximal beam current in the Linac will be 5 mA for maximal accelerating voltages of 1.0 MV and 2.3 MV for low and high betas respectively. SC cavities have typical loaded quality factor of $10^6$ and thus small resonant bandwidth (typically, in the order of 150 Hz), so that they are very sensible to perturbations presented in the next section. 
The SARAF-Linac LLRF system will also drive normal conducting cavities: one RFQ and three rebunchers. Their loaded quality factor is much smaller, typically 3500 for the rebuncher. 
%========================================
\section{Disturbance}
We generally consider two kind of disturbances according to their inherent properties: repetitive and non-repetitive\cite{Schmidt}. 

The repetitive components are predictable, it includes Lorentz force detuning and beam loading. Repetitive components can be corrected by feedforward and good synchronization of the components of the LINAC. However, we chose to not implement any feedforward functionnality, considering that the LLRF system would be fast enough to stabilize the field. This will be discussed later in the paper.

Non-repetitive disturbances on the other hand cannot be foreseen, it includes detuning due to the helium bath pressure variation and vibrations of the cavities. As the normal conducting cavity has large bandwidth, these disturbances could be neglected or compensated by a frequency tuner. However, the SC cavity is very sensitive to the helium bath pressure variation. 
\subsection{Instabilities of helium bath pressure}
SC cavities are very sensitive to helium bath pressure, leading to small oscillations changing the cavity geometry. Accelerator components, such as cooling systems and vacuum pumps, are sources of instabilities of the helium bath.

Typically, the pressure of the helium bath for SC cavity cooling could have a variation of $\pm 5$ mbar and the resonance frequency of the cavity has a sensibility to the bath lower than 5 Hz/mbar. So that, the helium bath leads to a resonance frequency variation of $\pm 25$ Hz. This small resonance frequency change has a significant impact on the field, as the mistuning leads to higher power reflection and phase shift. 

Considering the 100 Hz bandwidth of the SC cavity, the frequency variation could lead to an amplitude transmission lowered by 3\% and a phase shift of $14^\circ$.  

\subsection{Beam loading}

For beam loading simulations, we considered an amplitude of 100\%/11\% (of the accelerating field) and a phase of 160$^\circ$/90$^\circ$ for SC cavities and rebunchers respectively.

\section{Requirement}
Requirements for amplitude and phase field control were determined by the beam dynamic study: 1\% and 1$^\circ$, respectively. Considering the error during calibration of the system and the other sources of instability during operation (amplifiers, cables, etc.), the requirements for the LLRF regulation are: amplitude stability $< 0.1 \%$ and phase stability $ < 0.1 ^\circ$.

Considering the error caused by the vibration, we can estimate that the PID controller should have a proportional gain of at least 140 (which is the ratio between the expected phase shift of 14$^\circ$ and the required phase stability of 0.1$^\circ$). In addition, the measurement precision of the LLRF system should be below 0.1\% in amplitude and 0.1$^\circ$ in phase to achieve the beam dynamics requirements, such as the LLRF system only contributes to 20\% of the acceptable amplitude and phase error.

\begin{table}[!hbt]
    \centering
        \caption{Expected Performances of Superconducting Cavities with coupler $Q_{ext}$ of $10^6$ \cite{coupler}}.
    \begin{tabular}{p{3.5cm}|p{1.8cm}|p{1.8cm}}
    \hline
                    & Low $beta$ & High $beta$  \\
                      \hline
       $\beta{opt}$ & 0.091 & 0.181 \\ 
       Design $E_{acc} (MV/m)$ & 7 & 8.1 \\
       $E_{pk_max} (MV/m)$ & 34.5 &35.8 \\ 
       Target $Q_0$ @ 4.45K & $8\cdot10^8$ & $1.2\cdot 10^9$\\ 
       R/Q @ $\beta_{opt} (\Omega)$ &189 & 280\\ 
       Stored energy (J) & 5.7 &16.8\\
       
       \hline
    \end{tabular}
    \label{tab:SCcavities}
\end{table}
% =========================================
\section{Modelling}
Under some hypothesis (cf.\cite{controlRF}), the maximal acceptable gain can be estimated by:
\begin{equation}
    T_d = \frac{Q_L}{\pi \times f_0} \cdot ln \frac{K_p}{K_p-1}
\end{equation}

\begin{figure}[!hbt]
            \centering
            \includegraphics[width=8cm]{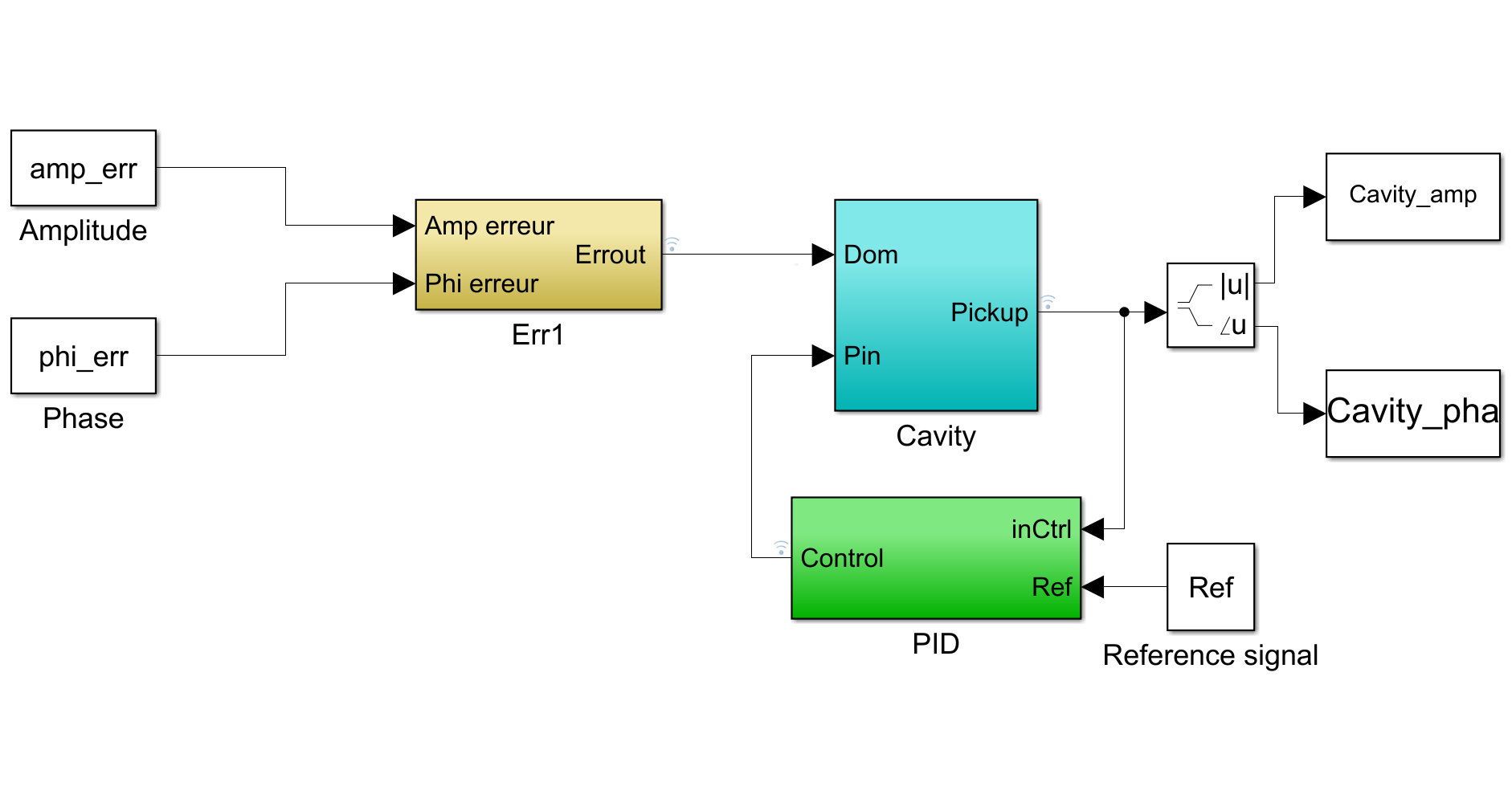}
            \caption{LLRF Simulink model.}
            \label{fig:llrf_model}
        \end{figure}

Where $T_d$ is the total delay of the system and $Q_L$ is loaded quality factor. In order to check the stability of the LLRF system with the estimated gain, we performed a simulation. Many RF control models have been developed\cite{luong, Schmidt}. Many perturbation source models can be easily inserted in the simulation carried out with Matlab/Simulink. Figure \ref{fig:llrf_model} illustrates the LLRF feedback loop model. Using this model, we can check the PID controller constituted by the following transfer function. 
        \begin{equation}
        \displaystyle u(t)=K_{\text{p}}\left(e(t)+{\frac {1}{T_{\text{i}}}}\int _{0}^{t}e(t')\,dt'+T_{\text{d}}{\frac {de(t)}{dt}}\right)
        \end{equation}
where, $K_p$ is the proportional gain, $K_p/T_i = Ki$ is integral gain and $K_pT_d=Kd$ is the derivative gain. For SARAF, we chose to set the derivative gain to 0. This configuration was already validated on Spiral 2 LLRF\cite{spiral}.
The mechanical vibration can be simplified modelled as a sinusoidal oscillation. 
        
For the simulations, we considered that the RF amplifier cannot generate more than 150\% of the nominal required power in operation. For this, the output RF amplitude was limited.
\subsection{Delay}
A time delay block is inserted into the signal path to simulate the time delay caused by the cables, signal processing and RF amplifier delay. 
According to the SARAF project configuration, some cables could have lengths up to 50 m. Considering all the cables required from the LLRF injection power output to the LLRF cavity power input, a time delay around 0.5 $\mu s$ was considered. According to the specifications of the LLRF system, its delay should be lower than 1 $\mu s$. In addition, the RF amplifier used for the SARAF project shows a typical time delay of 1 $\mu s$. For these reasons, we set the time delay in the simulation to 8 $\mu s$ with a large margin.

\section{Results}

\subsection{Estimation of the maximal acceptable gain}

Considering the previous equation, the estimated maximal acceptable gain is 226 for the SC cavities and 1.4 for rebunchers. We can observe, for the SC cavity that it is higher than the requirement of 140.

In the following sections, we will first focus on the SC cavities. We considered for simulations a slightly higher gain: 260.

\subsection{Startup of the SC cavity}

\begin{figure}
    \centering
    \includegraphics[width=8cm]{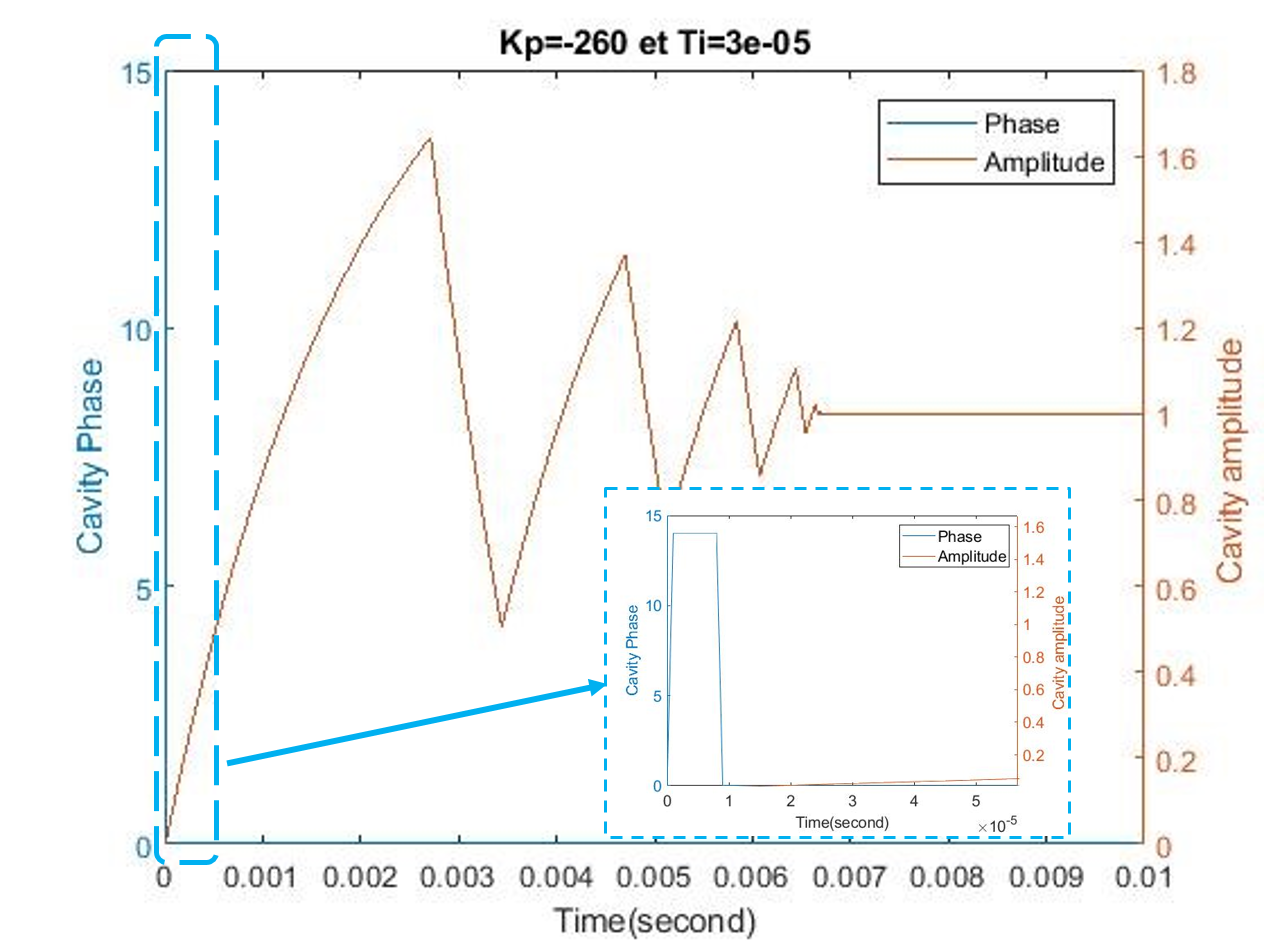}    
    \caption{Startup of a SC cavity with feedback}
    \label{fig:startup}
\end{figure}

Figure \ref{fig:startup} shows the behavior of the LLRF system when starting the cavity, in presence of the previously defined error. It shows that it requires about 7 ms to startup the cavity, without any feedforward. This is close to the typical response time of the cavity (inverse of the bandwidth). The cavity is very stable after this 7 ms.

This demonstrates that, without any feedforward, the cavity can be launched, and the beam can be inserted after about 7 ms. As the cavities will not be used in RF pulsed mode, this delay is not critical.

\subsection{Beam Loading in SC cavities}

\begin{figure}
    \centering
    \includegraphics[width =8cm]{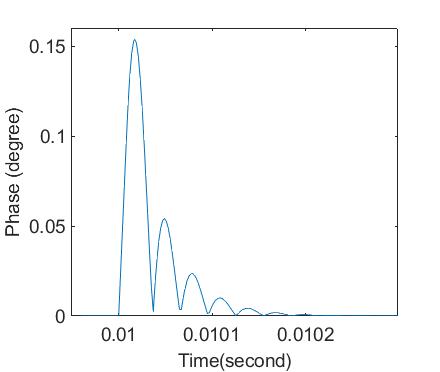}
    \caption{SC cavity phase variation when the beam arrives at 10 ms.}
    \label{fig:sc_beamloading}
\end{figure}

Figure \ref{fig:sc_beamloading} shows the effect of the beam loading in a SC cavity, modelled as a step signal arriving at 10 ms. Kp and Ti were defined to 260 and 30 $\mu s$ for this simulation. This simulation does not consider any feedforward, that is, synchronization between the LLRF system and the beam. Nevertheless, we observed the phase shift does not exceed 0.16$^\circ$, and does not require more than 10 $\mu s$ to get lower than 0.1$^\circ$. According to beam dynamics, this could be fast enough to have no impact on the operation.

This demonstrates that feedforward could be unnecessary even for compensating the beam loading.

\subsection{Beam Loading in rebunchers}
\begin{figure}
    \centering
    \includegraphics[width =8cm]{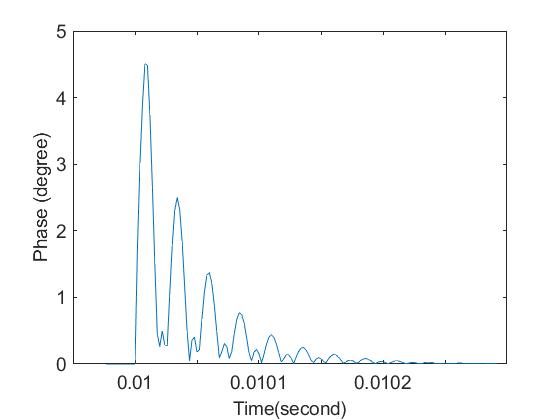}
    \caption{Rebuncher phase variation when the beam arrives at 10 ms.}
    \label{fig:rb_beamloading}
\end{figure}

Figure \ref{fig:rb_beamloading} shows the effect of the beam loading in a rebuncher. Kp and Ti were defined to 1.4 and 25 $\mu s$ for this simulation. According to this simulation, time to recover phase under 0.1$^\circ$ and amplitude under 0.1\% is about 150 $\mu s$.
Considering a time delay of 4 $\mu s$, a gain of 2.3 and a Ti of 20 $\mu s$, the time to recover is about 80 $\mu s$.

These cases have not been yet simulated with beam dynamics in order to understand whether it is critical or not in operation. If this is critical, feedforward would be necessary.

% ==============================================
\section{Schedule}
The public biding of LLRF system has closed in August 2019. The kick-off meeting will occur 1-2 weeks after the signature of CEA. The study and design phase will take 4 to 5 months. The first prototype LLRF system could be fabricated and delivered to CEA on the mid of 2020.  

At the same time, a test bench is under construction in the RF laboratory at CEA. For example, the necessary electronic devices, as RF amplifier and filter, and measurement instrument, as power meter, vector network analyzer. 
The first test of LLRF system could be performed with the fabricated rebuncher\cite{rebuncher} and the SC cavity \cite{cavity}. A RF power amplifier developed by SNRC has been tested with a water cooled $50\Omega$ load \cite{ampl}.
%================================================
\section{Conclusion}
The paper presents the cavity specification for SARAF project phase \uppercase\expandafter{\romannumeral2}. Using a simplified LLRF model, we investigated the effect of the disturbances, cavity startup and beam loading on the cavity field. The simulated results give us data to configure it and define if feedforward is necessary or not. The public biding of LLRF system for SARAF project is well passed. The test bench at laboratory is under preparation and the first prototype LLRF system will be ready for test on 2020.  
%=================================================
\section{Acknowledgement}
The authors acknowledge O.Piquet and M. Luong for discussions and support about their LLRF model, P. Nghiem and D. Uriot for information and experience about the beam dynamics.
%
% only for "biblatex"
%
\ifboolexpr{bool{jacowbiblatex}}%
	{\printbibliography}%
	{%
	% "biblatex" is not used, go the "manual" way

} % end \ifboolexpr
%
% for use as JACoW template the inclusion of the ANNEX parts have been commented out
% to generate the complete documentation please remove the "%" of the next two commands
% 
%%%\newpage

%%%\include{annexes-A4}

\end{document}